\documentclass[reprint,amsmath,amssymb,aps,prd,showpacs,twocolumn]{revtex4-1}

\usepackage{epsfig}
\usepackage{dcolumn}% Align table columns on decimal point
\usepackage{bm}% bold math
\usepackage{tikz}
\usepackage{graphicx, graphics, color}
\usepackage{dcolumn}% Align table columns on decimal point
\usepackage{bm}% bold math
\usepackage{hyperref}% add hypertext capabilities
\usepackage[mathlines]{lineno}% Enable numbering of text and display math
\newcommand{\be}{\begin{equation}}
\newcommand{\ee}{\end{equation}}
\newcommand{\ben}{\begin{eqnarray}}
\newcommand{\een}{\end{eqnarray}}

\newcommand{\cK}{{\cal K}}

\newcommand{\cO}{{\cal O}}

\newcommand{\cM}{{\cal M}}

\newcommand{\p}{\partial}
\newcommand{\na}{\nabla}

\newcommand{\Lie}{{\cal L}}

\newcommand{\tg}{\tilde g}

\newcommand{\ga}{\gamma}

\newcommand{\tR}{{\tilde R}}

\newcommand{\tna}{\tilde \na}

\begin{document}

\title{Uniqueness of higher-dimensional phantom field wormholes}
%%%%%%%%%%%%%%%%%%%%%%%%%%%%%%%%%%%%%%%%%%%%%%%%%%%%%%%%%%%%%%%%%%%%%%%%%%%%%%%%%%%%%%%%%%%%%%%%%%%%%%%%%%%%%
\author{Marek Rogatko} 
\email{rogat@kft.umcs.lublin.pl}
%,marek.rogatko@poczta.umcs.lublin.pl }
%\author{Karol I. Wysoki\'nski}
%\email{karol@tytan.umcs.lublin.pl}
\affiliation{Institute of Physics, % \protect \\
Maria Curie-Sklodowska University, % \protect \\
20-031 Lublin, pl.~Marii Curie-Sklodowskiej 1, Poland}

\date{\today}% It is always \today, today,

%%%%%%%%%%%%%%%%%%%%%%%%%%%%%%%%%%%%%%%%%%%%%%%%%%%%%%%%%%%%%%%%%%%%%%%%%%%%%%%%%%%%%%%%%%%%%%%%%%
\begin{abstract}
Based on the rigid positive energy theorem we proved the uniqueness of static spherically symmetric traversable wormholes with two asymptotically flat ends, being the higher-dimensional solutions
of Einstein scalar phantom field. The proof is valid under the auxiliary condition imposed on wormhole mass and scalar charge.
\end{abstract}

% 04.20.-q 	Classical general relativity
% 04.20.Fy 	Canonical formalism, Lagrangians, and variational principles
% 72.20.My 	Galvanomagnetic and other magnetotransport effects
\maketitle

%%%%%%%%%%%%%%%%%%%%%%%%%%%%%%%%%%%%%%%%%%%%%%%%%%%%%%%%%%%%%%%%%%%%%% 
\section{Introduction}
The existence of a transversable wormholes was proposed by Einstein and Rosen in 1935 (previously known as Einstein-Rosen bridges) \cite{ein35}. 
At the beginning of their history these objects were also considered as geometric models of elementary particles \cite{mis57}, as well as, to describe the initial data for Einstein equations \cite{mis60}.

Wormholes provide us the shortcuts through
spacetime by connecting two different far away spacetime points or even they supposed to join separate Universes \cite{mor88}. The object in question should not posses event horizon or physical singularities.
The simplest model for such an object, a throat which is held open by the presence phantom field, was presented in \cite{ell73}-\cite{ell79}. The choice of phantom field, the scalar field with a reserved sign in its kinetic 
energy term, is necessary in general relativity to furnish the required violation of the energy conditions, which is crucial in the case under consideration.
One should remark that contemporary astrophysical observations accommodate the evidence that due to the acceleration of our Universe, over sixty percent of its mass should constitute 
dark energy, being the ingredient which resembles phantom fields.
 
This idea was discussed in \cite{mor88amj,mor88}, where the motivation for the construction of wormhole in the realm of general relativity (GR) was given. On the other hand, in theories which authorize
the generalization of GR (higher curvature order, Gauss-Bonnet-dilaton gravity, see, e.g., \cite{kan11}-\cite{har13}), wormholes can be built with no use of such exotic kind of matter. 
Besides the static wormhole solutions the stationary axisymmetric ones also attracted the attention \cite{teo98}-\cite{kle14}.

As far as the observation implication of their existence is concerned, they were elaborated in different configurations, e.g., neutron star wormhole system \cite{dzh11,dzh14} or solely, wormholes which were 
responsible for gravitational lensing effects \cite{cra95,ned13}.  

Due to the possible way of building unification scheme, the subject of higher-dimensional wormhole physics was also paid attention to \cite{bro97,deb03}.
Among all in \cite{tor13} the generalization of Ellis wormhole solution was proposed.

The classification of these fascinating objects, in the context of the uniqueness theorem like for black hole, was initiated in \cite{rub89} and \cite{yaz17}. In \cite{rub89} the uniqueness theorem
for wormhole spaces with vanishing Ricci scalar was derived, while in \cite{yaz17} the uniqueness
of Ellis-Bronikov wormhole with phantom field was proposed.

In our paper we shall treat the case of higher-dimensional static spherically symmetric Ellis-type wormhole with two asymptotically flat ends, being subject to
Einstein scalar phantom field equations of motion. In what follows we shall use as a main tool conformal transformations and rigid positive energy theorem \cite{posen}. The first transformation
is used in order to check the outer boundary conditions. The other one was implemented to show the Ricci flatness, while the last conformal transformation enables us to show the conformal
flatness of the spacetime in question.

%%%%%%%%%%%%%%%%%%%%%%%%%%%%%%%%%%%%%%%%%%%%%%%%%%%%%%%%%%%%%%%%%%%
\section{Uniqueness theorem}
The action describing Einstein scalar phantom system is provided by
\be
S = \int d^n x \sqrt{-g}~ \Big( \frac{1}{2 \kappa_n^2} R + \frac{1}{2} \na_\alpha \phi \na^\alpha \phi \Big),
\ee
where $R$ is the Ricci scalar in $n$-dimensional spacetime, while $\na_\alpha$ denotes Levi-Civita connection in the manifold in question.
The energy momentum tensor for the exotic matter is given by
\be
T_{\alpha \beta} = - \na_{\alpha} \phi \na_{\beta} \phi + \frac{1}{2} g_{\alpha \beta} \na_\ga \phi \na^\ga \phi,
\ee
while the Einstein phantom matter equations of motion imply
\ben
G_{\alpha \beta} &=& \kappa_n^2~T_{\alpha \beta} (\phi),\\
\na_\delta \na^\delta \phi &=& 0.
\een
In our considerations we shall concentrate on the static spacetime in the strict sense, with a timelike Killing vector field $\xi_\alpha = (\p/\p t)_\alpha$ determined in each point of the manifold.
The line element of the considered spacetime can be written as  
\be
ds^2 = - N^2 dt^2 + g_{ij} dx^i dx^j,
\ee
where $g_{ij}$ certifies the metric tensor of $(n-1)$-dimensional Riemannian manifold, whereas $N$ is a smooth lapse function.
The lapse function and the metric tensor components are independent on time coordinate as the quantities defined on the hypersurface of constant time.

The dimensionally reduced equations of motion are provided by
\ben \label{rr}
{}^{(n-1)}R_{ij} = \frac{1}{N} {}^{(g)} \na_i {}^{(g)} \na_j N &-& \kappa_n^2 {}^{(g)} \na_i  \phi {}^{(g)} \na_j \phi,\\ \label{np}
{}^{(g)} \na_i \bigg( N~{}^{(g)} \na^i \phi \bigg) &=& 0, \\ \label{nn}
{}^{(g)} \na_i {}^{(g)} \na^i N &=& 0,
\een 
where ${}^{(n-1)}R_{ij}$ and ${}^{(g)} \na_i $ are Ricci scalar curvature and connection bounded with $(n-1)$-dimensional spacetime, respectively. 

The phantom scalar field will be subject to the condition $\Lie_\xi \phi = 0$, where $\Lie_\xi $ for the Lie derivative with respect to the timelike Killing vector field $\xi_\alpha$.\\
The above assumption about the strict static spacetime, ensures that we have no event horizons in the manifold in question. Moreover we assume that
the $(n-1)$-dimensional submanifold is complete, i.e., the $(n-1)$-dimensional hypersurfaces of constant time are singularity free.
For a compact subset $\cK \subset {}^{(n-1)}\cM$, consisting of two ends ${}^{(n-1)}\cM_{\pm}$ diffeomorphic to $R^{n-1}/B^{n-1}$, where $B^{n-1}$
is closed unit ball situated at the origin of $R^{n-1}$, one can find a standard coordinate system in which the
following expansion is proceeded:
\ben
g_{ij} &=& \bigg( 1 + \frac{2}{n-3} \frac{M_\pm}{r^{n-3}} \bigg) \delta_{ij} + \cO \Big(\frac{1}{r^{n-2}}\Big),\\
N &=& N_{\pm} \bigg(1 - \frac{M_\pm}{r^{n-3}} \bigg) +  \cO \Big(\frac{1}{r^{n-2}} \Big),\\
\phi &=& \phi_{\pm} - \frac{q_\pm}{(n-3) r^{n-3}} + \cO \Big(\frac{1}{r^{n-2}} \Big),
\een
where $N_\pm>0,~\phi_\pm,~\mu_\pm,~q_\pm$ are constant.
$M_\pm$ and $q_\pm$ represent the ADM masses up to a constant factor and charges of the two ends ${}^{(n-1)}\cM_{\pm}$, while $r^2 = x_m x^m$.

As in three-dimensional case, let us take into account $(n-1)$-dimensional metric tensor defined by the conformal transformation of the form
\be
{}^{(n-1)}\tg_{ij} = N^{\frac{2}{n-3}} ~g_{ij}.
\ee
Then the Ricci curvature tensor in the conformally rescaled metric yields
\ben  \label{rij} \nonumber
{}^{(n-1)} \tR(\tg)_{ij} &=& \frac{1}{N^2} \Big( \frac{n-2}{n-3} \Big) {}^{(n-1)}\tna_i N {}^{(n-1)}\tna_j N \\ 
&-& \kappa_n^2 {}^{(n-1)}\tna_i \phi {}^{(n-1)}\tna_j \phi,
\een
and for the lapse and phantom field, one obtains the following relations:
\ben
{}^{(n-1)}\tna_i {}^{(n-1)}\tna^i N &=& 0,\\
{}^{(n-1)}\tna_i {}^{(n-1)}\tna^i \phi &=& 0.
\een
In order to rewrite the equation (\ref{rij}) as a function of the phantom field, we shall search for the relation of the lapse function and the scalar phantom field.
But before proceeding to this subject we define the scalar field mass and charge. Namely, having in mind the fact that $N$ is a harmonic function in ${}^{(n-1)}\cM $,
one gets
\be
\int_{{}^{(n-1)}\cM } {}^{(g)} \na_i {}^{(g)} \na^i N = \int_{S^{(n-2)}_\infty}{}^{(g)} \na_i N - \int_{\Sigma} {}^{(g)} \na_i  N,
\ee
where $S^{(n-2)}$ is an $(n-2)$-dimensional sphere at infinity, $\Sigma$ is a hypersurface.
This relation enables us to define mass in the manifold with $g$-metric tensor
\be
M = \int_{S^{(n-2)}_\infty}{}^{(g)} \na_i N=\int_{\Sigma_{wh}} {}^{(g)} \na_i  N.
\ee
The same procedure can be undertaken in order to define phantom charge, i.e., using the equation (\ref{np}) and the asymptotic behavior of $N,~\phi,~g_{ij}$, we arrive at
\be
q= \int_{S^{(n-2)}_\infty}{}^{(g)} \na_i \phi =\int_{\Sigma_{wh}} {}^{(g)} N_0 \na_i  \phi.
\ee
Like in Ref.\cite{yaz15}, one can consider 
$J_{a}  = \phi~{}^{(g)} \na N - N~\ln N {}^{(g)} \na \phi$ which is the conserved quantity in ${}^{(n-1)}\cM $, i.e., ${}^{(g)} \na_m J^m =0$.
The same reasoning as in the four-dimensional case leads to the conclusion that
\be
\phi = \frac{q}{(n-3)~M}~\ln N.
\ee
Thus, the above relation leads to the relation binding conformally rescaled Ricci tensor with scalar phantom field. It implies
\be
{}^{(n-1)} \tR(\tg)_{ab} = - (n-2)~\Big[ \frac{\kappa_n^2}{n-2} - \frac{(n-3) M^2}{q^2} \Big] {}^{(n-1)}\tna_a \phi {}^{(n-1)}\tna_b \phi,
\ee
If we redefine the scalar field $\phi$ in the form as follows:
\be
\theta = \sqrt{\frac{\kappa_n^2}{n-2} - \frac{(n-3) M^2 }{q^2}} ~\phi,
\ee
one achieves the underlying equations of motion provided by
\ben
{}^{(n-1)} \tR(\tg)_{ab} = &-& (n-2)~{}^{(n-1)}\tna_a \theta {}^{(n-1)}\tna_b \theta,\\
{}^{(n-1)}\tna_a {}^{(n-1)}\tna_a \theta &=& 0.
\een
The $(n-1)$-dimensional Riemannian manifold $({}^{(n-1)}\cM, ~{}^{(n-1)}\tg_{ij} )$ is a complete asymptotically flat manifold with two ends ${}^{(n-1)}\cM_{\pm}$ of vanishing masses.
As in three-dimensional case the fact that ${}^{(n-1)}\tg_{ij} $ is complete follows from its definition, completeness of $g_{ij}$ and the inequality $N_-<N<N_+$. On the other hand,
the condition of vanishing the total mass of each ends stems from the asymptotic behavior of  ${}^{(n-1)}\tg_{ij} = \delta_{ij} + \cO(1/r^{n-2})$, as well as, the asymptotical conditions imposed on
$g_{ij}$ and the lapse function $N$.

One has also that the following inequality is provided
\be
M^2 \ge \frac{\kappa_n^2~q^2}{(n-2)(n-3)},
\ee
the form of which is motivated by the fact that the other form of it will pose the contradiction with the positive energy theorem \cite{posen}.

In the next step, let us define the conformally rescaled metric tensor $\ga_{ij}$, which yields
\be
\ga_{ij} = \Omega^2~ \tg_{ij},
\ee
where the conformal factor implies
\be
\Omega = \bigg( \sin^2 \Big( \frac{\sqrt{n-3}~\theta}{2} \Big) \bigg)^{\frac{1}{n-3}}.
\ee
By the direct calculation one can show that the Ricci scalar curvature for the metric $\ga_{ij}$ fulfills
\be
R(\ga_{ab}) = 0.
\ee
As in \cite{bun87}, the ${}^{(n-1)}\cM_-$ end can be compactified by adding the point at infinity.
By construction the resulting manifold ${}^{(n-1)}{\tilde {\cM}}=
{}^{(n-1)}\cM \cup \{ \infty \}$ is geodesically complete with one end ${}^{(n-1)}\cM_+$. Moreover,
the Ricci scalar $R(\ga)$ vanishes on it. The form of the $(n-1)$-dimensional metric tensor yields that the total mass also vanishes on
$ {}^{(n-1)}{\tilde {\cM}}$. As a consequence of the rigid positive energy theorem \cite{posen}, the manifold in question must be flat, i.e., 
it is isometric to $(R^{(n-1)},~\delta_{ij})$ manifold. On the other hand, the metric tensor $h_{ij}$ and ${}^{(n-1)}g_{ij}$ are conformally flat.
In other words ${}^{(n-1)}g_{ij}$ can be provided by the relation
\be
{}^{(n-1)}g_{ij} = v^{\frac{2}{n-3}}~\delta_{ij},
\ee
where $v = 1/N$. The $R_{00}$ component of the Einstein-phantom equations reduces to
\be
{}^{(\delta)} \na_j {}^{(\delta)} \na^j v = 0,
\ee
where we denoted by ${}^{(\delta)} \na_j $ the flat connection.

Because of the fact that the lapse function is a harmonic one, we can implement it to define a local coordinates in the  neighborhood $\cK \subset  {}^{(n-1)}{\tilde {\cM}}$.
Thus, the metric tensor on ${}^{(n-1)}{\tilde {\cM}}$ can be given by
\be
\delta_{ij}~dx^i dx^j = \rho^2~dN^2 + h_{AB}~dx^A dx^B,
\ee
where $\rho^2 = \na_k N \na^k N$.

The ${}^{(n-1)}{\tilde {\cM}}$ manifold is totally geodesic in the sense that any of its sub-manifold geodesic is also  a geodesic in the manifold in question.
On the other hand, the embedding of ${}^{(n-1)}{\tilde {\cM}}$ into Euclidean $(n-1)$-manifold is totally umbilical \cite{kob69},
which results that each connected component of ${}^{(n-1)}{\tilde {\cM}}$ is a geometric sphere with a certain radius. The considered embedding is also rigid \cite{kob69}, i.e.,
that one can always locate one connected wormhole of a certain radius $\rho$, at $r=r_0$ surface on ${}^{(n-1)}{\tilde {\cM}}$, without loss of generality.
Thus we arrive at a boundary value problem for the Laplace equation on the base space $\Theta = E^{n-1}/B^{n-1}$, with a Dirichlet boundary conditions.
The system in question is characterized by a parameter which fixes the radius of the inner boundary  and constitutes wormhole of a specific radius $\rho$. The asymptotic decay condition for $v$
is given by $v = 1 + \cO(r^{n-3})$, with the demand that $r \rightarrow \infty$. Let us suppose that we have two solutions subject to the same boundary value problem. Having in mind the Green
identity and integrating over the volume $\Theta$, we arrive at the expression
\ben \nonumber
\Big( \int_{r \rightarrow \infty} - \int_{\Sigma_{wh}} \Big) (v_1-v_2)~\frac{\p}{\p r} (v_1-v_2) dS \\
= \int_\Theta \mid \na (v_1 - v_2) \mid^2 d \Theta.
\een
Having in mind that the left-hand side vanishes, we obtain the conclusion that the two solutions have to be identical.

All the above lead to the main conclusion that:\\
the only static spherically symmetric asymptotically flat tranversable wormhole spacetime, being the solution of $n$-dimensional Einstein-phantom scalar field
equations of motion, whose mass and scalar charge are subject to the auxiliary condition
\be
M^2 \ge \frac{\kappa_n^2~q^2}{(n-2)(n-3)}, \nonumber
\ee
is the Torii-Shinkai higher-dimensional wormhole.

%%%%%%%%%%%%%%%%%%%%%%%%%%%%%%%%%%%%%%%%%%%%%%%%%%%%%%%%%%%%%%%%%%%%%%%
\section{Conclusions}
In our paper we considered $n$-dimensional static spherically symmetric Ellis-type wormhole with two asymptotically flat ends. By assuming that the mass
and the scalar charged of phantom field are subject to the inequality, implementing the rigid positive energy theorem, we show the uniqueness of such solution of Einstein scalar phantom theory.

As a future investigation it will be interesting to consider higher-dimensional wormhole solution with $U(1)$-gauge fields or higher curvature correction of the model.
We hope to return to these problems elsewhere.

%%%%%%%%%%%%%%%%%%%%%%%%%%%%%%%%%%%%%%%%%%%%%%%%%%%%%%
\begin{acknowledgments}
 {MR was partially supported by the grant DEC-2014/15/B/ST2/00089 of the National Science Center.}
% and KIW by the grant \mbox{DEC-2014/13/B/ST3/04451}.
 \end{acknowledgments}

%%%%%%%%%%%%%%%%%%%%%%%%%%%%%%%%%%%
%\end{document}
%%%%%%%%%%%%%%%%%%%%%%%%%%%%%%%%%%%%%%%%%%%%%%%%%%%%%%%%%%%%%%%%%%%%%%%%%%%%%%%%%
%%%%%%%%%%%%%%%%%%%%%%%%%%%%%%%%%%%%%%%%%%%%%%%%%%%%%%%%%%%%%%%%%%%

%%%%%%%%%%%%%%%%%%%%%%%%%%%%%%%%%%%%%%%%%%%%%%%%%%%%%%%%%%%%%%%%%%%%%%%%%%%%%%%%%%%%%%%%%%%%%%%%%%%%%%%%%%%%%%%%%%%%

\end{document}